\documentstyle[preprint,prl,aps,eqsecnum]{revtex}

\begin{document}
\draft
\preprint{\vbox{
\hbox{TRI-PP-94-1}
\hbox{VPI-IHEP-94-1}
\hbox{February, 1994}
\hbox{hep-ph/9402259}
}}

\title{
Phenomenological Consequences of Singlet Neutrinos
}
\author{Lay Nam Chang\thanks{Currently at Physics Division,
National Science Foundation, Arlington, Virginia 22230, USA, under a
contract with the National BioSystems, Rockville, Maryland 20852, USA.}}
\address{Institute for High Energy Physics\\
Virginia Polytechnic Institute and State University\\
Blacksburg, Virginia 24061-0435, U.S.A.}

\author{Daniel Ng and John N. Ng}
\address{
TRIUMF, 4004 Wesbrook Mall\\
Vancouver, B.C., V6T 2A3, Canada
}
\maketitle
\begin{abstract}
{
In this paper, we study the phenomenology of right-handed
neutrino isosinglets.  We consider the general situation where
the neutrino masses are not necessarily given by
$m_D^2/M$, where
$m_D$ and $M$ are the Dirac and Majorana mass terms respectively.
The consequent
mixing between the light and heavy neutrinos is then not suppressed, and
we treat it as an independent parameter in the analysis.
It turns out that
$\mu-e$ conversion is an important experiment in placing limits on the
heavy mass scale ($M$) and the mixing.
Mixings among light neutrinos are constrained by
neutrinoless double beta decay, as well as by solar and atmospheric
neutrino experiments.
Detailed one-loop calculations for lepton number violating vertices are
provided.
}
\end{abstract}
\pacs{PACS numbers: 12.15.-y, 13.10.+q, 13.15.-f, 14.60.Ef, 14.60.Gh}

\section{Introduction}
\label{sec:intro}

There is no direct evidence so far that neutrinos have mass.   Indirectly,
however, measurements on solar neutrino fluxes suggest that indeed they
do have
masses, albeit at values which are considerably smaller than those
for charged fermions\cite{solar}.   Within the standard
model, this situation is accomodated quite naturally
by restricting Higgs fields to
the usual isodoublets, so that there are no direct Yukawa couplings
among  the left-handed lepton fields and scalar bosons.    Nevertheless,
gravity effects could induce a dimension five operator, but these
would imply Majorana neutrino masses of the order $m_\nu \sim v^2/M_{Pl}
\sim 10^{-5}~eV$, where $v=250~{\rm GeV}$ is the scale of electroweak breaking
and $M_{Pl}=10^{19}~{\rm GeV}$ is the Planck mass.
In what follows, we will ignore such contributions.

Generating neutrino masses poses somewhat different problems from
those for charged fermions.   This is primarily because neutral fermions
could acquire Majorana masses, and so the whole question of mixing
angles and their attendant CP phases needs to be re-examined\cite{shrock}.
The most elementary way of generating neutrino masses
would be through the introduction of neutral electroweak singlet fermion
fields into the theory.
Detecting finite masses for neutrinos therefore would provide
a direct way for probing structure and dynamics beyond those of
the standard model.

Right-handed neutrinos, which are electroweak singlet fermions,
can have gauge invariant Majorana masses, $M$.
The presence of a Higgs isodoublet induces
Yukawa couplings of left- and right-handed neutrinos.  Thus, left- and
right-handed neutrinos are linked together by Dirac masses, $m_D$.
The left-handed neutrinos acquire their Majorana masses, which is given
by $m_\nu~=~m_D^2/M$, when we integrate out the heavy right-handed
neutrinos.  This is called ``see-saw'' mechanism\cite{seesaw}.
The mixing of the left- to the right-handed neutrinos, given by
$m_D/M$, can be rewritten as $\sqrt{m_\nu/M}$.  As a result, exotic
processes, such
as $\mu \rightarrow e \gamma$, $\mu \rightarrow 3 e$ and $\mu-e$
conversion in nuclei, are very suppressed by the smallness of light
neutrino masses.

In the analysis to be presented below,
we consider the situation where the light
neutrinos are not given by $m_D^2/M$.  This is possible when there are
more than one right-handed neutrino.  Hence, the mixing, $m_D/M$,
will be independent of the light neutrino masses.
Within such a context, it will be sufficient for three generations
of left-handed neutrinos and an additional right-handed neutrino field
$\nu^c$ to
illustrate the kinds of bounds on neutrino masses and mixings
that can be extracted from existing data.
This model can be considered as a remnant
of some higher energy theory manifested at the current low
energy scale, and $\nu^c$ as an effective collection of arbitrary number
of right-handed neutrino fields.

The presence of  $\nu^c$ can give rise to much interesting phenomenology.
In addition to neutrino masses and mixings, there can be lepton
family number violating processes,
violations of generation
universality, and off diagonal neutral current couplings.
In this paper, we will consider this phenomenology in detail,
and examine how available data constrain the parameters in this scenario.
CP violation will not be considered here.
We first formulate the model in Sec.~\ref{sec:model}.
Constraints of the model, obtained from $Z$ decays and
universalities in charged current processes, are given in
Sec.~\ref{sec:constraints}.
In Sec.~\ref{sec:rare} and~\ref{sec:nonudecay},
we use the information obtained in Sec. III to
calculate lepton family number violating processes and neutrinoless
double beta decay.
In Sec.~\ref{sec:oscillation}, we discuss neutrino oscillations.
Finally, we will conclude our analysis in Sec.~\ref{sec:con}.
Although the model we are studying is not new, to the best of our
knowledge, the idea of relaxing the see-saw mass
relationship has not been studied.  In addition, the
detailed results on rare decays have not been presented before.

\section{formulation of the one singlet model}
\label{sec:model}

When one $\nu^c$ is added to the standard model, the new Yukawa interactions
that must be included are given by
\begin{equation}
\label{yukawa}
{\cal L}_Y(\nu^c) = -\frac{g}{\sqrt{2}m_W} \sum_{\alpha=e,\mu,\tau}
a_\alpha\pmatrix{\overline{\nu_{\alpha}} & \overline{\alpha_L}}
\pmatrix{ \frac{1}{\sqrt{2}}(H^0-iG^0) \cr - G^-} {\overline{\nu^c}} + h.c. \ ,
\end{equation}
where $a$'s are assumed to be real.  Without loosing any generality,
we can define the charged leptons $\alpha_L$ to be given by their mass
eigenstates.  Since $\nu^c$ is a gauge singlet of the standard model,
it can pick up a Majorana mass,
\begin{equation}
\label{MR}
{\cal L}_{mass}(\nu^c) = - \frac{1}{2}M \nu^c~ \nu^c + h.c. \ .
\end{equation}
$\nu_\alpha$ and $\nu^c$ are the two-component Weyl fields.
When the ${\rm SU(2)\times U(1)}$ gauge symmetry of the standard model
is broken spontaneously, mixings among the gauge eigenstates
$\nu_\alpha$ and $\nu^c$ are induced, leading to the following
mass matrix:
\begin{equation}
\label{massterm}
\frac{1}{2}\pmatrix{\nu_{e} & \nu_{\mu} & \nu_{\tau} & \nu^c} {\cal M}
 \pmatrix{\nu_{e} \cr \nu_{\mu} \cr \nu_{\tau} \cr \nu^c} + h.c. \ ,
\end{equation}
where
\begin{equation}
\label{massmatrix}
{\cal M}=
\pmatrix{0&0&0&a_e\cr0&0&0&a_\mu\cr0&0&0&a_\tau\cr a_e&a_\mu&a_\tau&M}\; .
\end{equation}
${\cal M}$ can be diagonalized by a rotational matrix $\cal O$,
\begin{equation}
\label{diagonal}
{\cal O}^T {\cal M} {\cal O} =
\pmatrix{0&0&0&0 \cr 0&0&0&0 \cr 0&0&m_3&0 \cr 0&0&0&m_4}\ .
\end{equation}
$\cal O$, defined as $\nu_\alpha=\sum_{i=1}^4{\cal O}_{\alpha i} \nu_i$
($\alpha=e$, $\mu$, $\tau$ and $R$), is explicitly given by
\begin{equation}
\label{O}
{\cal O}=\pmatrix{c_1&s_1c_2&s_1s_2c_3&s_1s_2s_3\cr
-s_1&c_1c_2&c_1s_2c_3&c_1s_2s_3 \cr 0&-s_2&c_2c_3&c_2s_3 \cr 0&0&-s_3&c_3}
\pmatrix{1&0&0&0 \cr 0&1&0&0 \cr 0&0&i&0 \cr 0&0&0&1} \ ,
\end{equation}
where we adopt the abbreviation $s_i=\sin\theta_i$ and $c_i=\cos\theta_i$.

Eq.~(\ref{O}) is defined in a way that both eigenmasses, $m_3$ and $m_4$, are
positive.   The `$i$' in Eq.~(\ref{O}) indicates that
$\nu_3$ and $\nu_4$ have opposite CP tansformation.  The mixing
angles among the light neutrinos are given by
\begin{equation}
\label{mixlight}
s_1=\frac{a_e}{\sqrt{a_e^2+a_\mu^2}} \ ;
s_2=\frac{\sqrt{a_e^2+a_\mu^2}}{\sqrt{a_e^2+a_\mu^2+a_\tau^2}} \ ;
\end{equation}
whereas the mixing between the light and heavy neutrinos is
\begin{equation}
\label{mixing}
s_3^2= \frac{m_3}{m_3+m_4} \ .
\end{equation}
The masses for the two massive neutrinos are given as
\begin{equation}
\label{m3}
m_3=\frac{-M+\sqrt{M^2+4(a_e^2+a_\mu^2+a_\tau^2)}}{2}~
\end{equation}
\begin{equation}
\label{m4}
m_4=\frac{M+\sqrt{M^2+4(a_e^2+a_\mu^2+a_\tau^2)}}{2} \ .
\end{equation}
The diagonalization condition, Eq.~(\ref{diagonal}), are used when
we calculate the $Z$ penguin diagrams, see Appendix~\ref{app:one-loop}.
Note thate for $M^2 \gg (a_e^2+a_\mu^2+a_\tau^2)$, we have the see-saw
mass for $\nu_3$, $m_3\sim(a_e^2+a_\mu^2+a_\tau^2)/M$.

Notice that $s_3$ is suppressed by the square root of the ratio
of light to heavy neutrino masses, in accordance with the
general arguments presented in the introduction.
As we have already pointed out there, to avoid such a suppression,
one requires more than one right-handed neutrino state.
When there are more than one right-handed neutrinos, see-saw
relationships, Eq.~(\ref{mixing}) and (\ref{m3}), do not necessarily hold.
We furnish details on how this can come about in appendix~\ref{example}.
In what follows, we shall accomodate such an eventuality by treating
$s_3$ as an independent parameter and
continue to consider $\nu^c$  as an effective collection of arbitrary
number of right-handed neutrinos.   This scenario may well be remnants
of symmetries which are manifest at higher energies.

The consequent charged
current interactions of $W$ gauge boson, in four component notation,
are given by
\begin{equation}
\label{Wen}
{\cal L}_W = \frac{g}{\sqrt{2}}W^-_\mu
\sum_{\alpha=e,\mu,\tau}\sum_{i=1,..,4}{\cal O}_{\alpha i}~
\overline{\alpha}\ \gamma^\mu \frac{1-\gamma_5}{2}\ \nu_i + h.c \ .
\end{equation}
The neutral current interactions of $Z$ gauge boson can be obtained
straightforwardly.   The interaction
remains the same as in the standard model for the charged leptons and
remains flavor diagonal.  However, there will be
flavor changing pieces induced by $\nu^c$.  The interactions in four
component notation are given
by
\begin{eqnarray}
\label{Zee}
{\cal L}_{Z\bar e e} &=&\frac{g}{\cos\theta_W} Z_\mu \sum_{\alpha=e,\mu,\tau}
{\overline \alpha}\ \gamma^\mu \left[
g_L \frac{1-\gamma_5}{2}+g_R \frac{1+\gamma_5}{2}\right]\alpha \\
\label{Znn}
{\cal L}_{Z\bar \nu \nu}&=&\frac{g}{4\cos\theta_W}
   Z_\mu \sum_{i,j=1}^{4} {\overline\nu_i}\ \gamma^\mu \left[
L_{ij} \frac{1-\gamma_5}{2}+R_{ij} \frac{1+\gamma_5}{2}\right] \nu_j \ ,
\end{eqnarray}
where
\begin{mathletters}
\begin{eqnarray}
g_L &=& -\frac{1}{2}+\sin^2\theta_W \ , \\
g_R &=& \sin^2\theta_W \ ,
\end{eqnarray}
\end{mathletters}
and
\begin{mathletters}
\begin{eqnarray}
L_{ij}&=&\delta_{ij}-{\cal O}^\ast_{Ri}{\cal O}_{Rj} \ , \\
R_{ij}&=&-\delta_{ij}+{\cal O}^\ast_{Rj}{\cal O}_{Ri}
\end{eqnarray}
\end{mathletters}

The interactions involving Goldstone bosons ($G^\pm$, $G^0$) and
the physical Higgs scalar ($H^0$) can be obtained from Eq.~(\ref{yukawa}),
and, with the help of Eqs.~(\ref{diagonal}), are given by
\begin{eqnarray}
\label{G}
{\cal L}_{G^-} &=& \frac{g}{\sqrt2m_W}G^-
\sum_{\alpha=e,\mu,\tau}\sum_{i=1,..,4}{\cal O}_{\alpha i}
{}~m_i~{\overline \alpha}~\frac{1+\gamma_5}{2}~\nu_i \; ,  \\
\label{G0}
{\cal L}_{G^0}&=&\frac{ig}{4m_W}G^0 \left[
\sum_{i=1,..,4}m_i~\nu_i^TC\gamma_5\nu_i
+ \sum_{i,j=1,..,4}M\left({\cal O}_{R i}{\cal O}_{R j}
{}~\nu_i^TC\frac{1-\gamma_5}{2}\nu_j
-h.c.\right) \right] \; , \\
\label{H0}
{\cal L}_{H^0} &=&-\frac{g}{4m_W}H^0\left[
\sum_{i=1,..,4}m_i~\nu_i^TC\nu_i
- \sum_{i,j=1,..,4}M\left({\cal O}_{R i}{\cal O}_{Rj}~\nu_i^TC
\frac{1-\gamma_5}{2}\nu_j + h.c.\right)\right] \; .
\end{eqnarray}

{}From Eqs.~(\ref{Znn}) and (\ref{H0}),  we show explicitly the flavor
changing coupling induced by $\nu^c$.  Using Eq.~(\ref{O}),
Eqs.~(\ref{Znn}), (\ref{G0}) and (\ref{H0}) can be rewritten as
\begin{eqnarray}
\label{Znn1}
{\cal L}_{Z\bar \nu \nu}&=&-\frac{g}{4\cos\theta_W} Z_\mu \left[
\sum_{i=1,2}{\overline\nu_i}\gamma^\mu\gamma_5\nu_i
+c_3^2{\overline \nu_3}\gamma^\mu\gamma_5\nu_3
+s_3^2{\overline \nu_4}\gamma^\mu\gamma_5\nu_4
+2is_3c_3{\overline \nu_3}\gamma^\mu\nu_4\right] \ , \\
\label{G01}
{\cal L}_{G^0} &=& \frac{ig}{4m_W}G^0 \left[
\sum_{i=1,..,4}m_i~\nu_i^TC\gamma_5\nu_i
+M\left(s_3^2\nu_3^TC\gamma_5\nu_3+c_3^2\nu_4^TC\gamma_5\nu_4
-2ic_3s_3\nu_3^TC\nu_4\right)\right] \ , \\
\noalign{\vbox{\vskip\abovedisplayskip \hbox{and}\vskip\belowdisplayskip}}
\label{H01}
{\cal L}_{H^0} &=&-\frac{g}{4m_W}H^0\left[
\sum_{i=1,..,4}m_i~\nu_i^TC\nu_i
+M\left(s_3^2\nu_3^TC\nu_3+c_3^2\nu_4^TC\nu_4-2ic_3s_3\nu_3^TC\gamma_5\nu_4
\right)\right] \ ,
\end{eqnarray}
respectively.   Since $\nu_3$ and $\nu_4$ have opposite CP because of
the structure of the mass matrix we have assumed, the flavor
changing interactions in Eqs~(\ref{Znn1}) and (\ref{H01})
have different Lorentz structure from that of the flavor conserving terms.
Hence, the flavor changing decays of $Z$ and $H^0$
may offer new channels to search for the existence of a right-handed
neutrino.

\section{constraints of the model}
\label{sec:constraints}

One of the predictions of this model is that two of the neutrinos
($\nu_{1,2}$) are massless at tree level.  These two massless neutrinos,
which are not protected by symmetries, will pick up Majorana masses
at  higher order loops\cite{babuma}, but their eventual masses
are negligibly small.  For our purpose, we simply assume
these two neutrinos to be massless.  The other two neutrinos ($\nu_{3,4}$)
are massive; we define $m_3 \leq m_4$.  The decays of $\nu_4$ present
us with rich class of phenomena.
To avoid any conflict with the
cosmological and astrophysical constraints \cite{bouquet},  we take
$m_4$ to be greater than $O(1)~\rm GeV$.

As seen in Eq.~(\ref{Znn1}), the presence of a right-handed singlet
 induces  flavor changing neutral currents among neutrinos.
In addition, the strength of $Z-\nu_3-\nu_3$ coupling
is reduced by a factor of $c_3^2$ relative to $Z-\nu_1-\nu_1$ and
$Z-\nu_2-\nu_2$.  Therefore, the invisible width of $Z$ gauge boson will
provide a stringent limit on the mixing parameters $s_3$.
If $\nu_4$ is heavier than $Z$,
 $s_3$ can be constrained from the invisible width of $Z$ gauge
boson.  A standard calculation using Eq.~(\ref{Znn1}) modifies the
formula for the number of light neutrino species as measured by LEP
\begin{equation}
\label{Nn}
N_\nu=2+(1-s_3^2)^2 \ .
\end{equation}
At $90\%$ C.L., $N_\nu$ is greater than $2.95$ \cite{langacker}, leading
to
\begin{equation}
\label{s3bound}
{s_3}^2 \leq 2.69\times10^{-2} \ .
\end{equation}
If $\nu_4$ is lighter than $Z$, the decays
$Z\rightarrow \nu_3~\nu_4$ or $ \nu_4~\nu_4$ are allowed.  If $m_4$ is
heavier than $O(1)~\rm GeV$, $\nu_4$ will decay within detectors, leaving
exotic signatures such as $Z\rightarrow e~\mu + X$.  Recent experimental
results on the search for lepton flavor violation in Z decays can be found in
Ref.~\cite{L3}.  The absence of
these exotic signatures then provides a very stringent constraint on
$s_3$.   Since the decays of $\nu_4$ are so numerous, we use a
conservative bound of
\begin{equation}
\label{branching}
B(Z\rightarrow \nu_3 \nu_4,~\nu_4 \nu_4 ) \leq 1\times10^{-5}
\end{equation}
to constrain $s_3$.  Combining Eqs.~(\ref{Nn}) and (\ref{branching}), we
plot the upper bound on $s_3$ as a function of $m_4$ in Fig.
\ref{fig:s3}.

Since $\nu_4$ is not kinematically allowed for the muon decay $\mu
\rightarrow e \nu \nu$,  only the first three neutrinos play a role.
The Fermi coupling constant $G_F$ extracted from muon lifetime is given
by
\begin{equation}
\left(\frac{G_F}{\sqrt{2}}\right)^2=\left(\frac{g^2}{8m_W^2}\right)^2
\left[1-|{\cal O}_{e4}|^2\right]\left[1-|{\cal O}_{\mu 4}|^2\right]
\end{equation}
When radiative corrections are included in the on shell scheme
\cite{sirlin}, the precisely measured quantity $m_W/m_Z$ can be related
to $G_F$ in the following way,
\begin{equation}
1-\frac{m_W^2}{m_Z^2}=\frac{A_0^2}{m_w}
\left[1-|{\cal O}_{e4}|^2\right]^{1/2}\left[1-|{\cal O}_{\mu 4}|^2\right]^{1/2}
\frac{1}{1-\Delta r} \ ,
\end{equation}
where $A_0^2=\pi\alpha_{em}/\sqrt{2}/G_F=(37.2803~{\rm GeV})^2$.  The
quantity $\Delta r$ depends on the masses of top quark and Higgs.
Taking $100~{\rm GeV} \leq m_H \leq 1~{\rm TeV}$ and $100~{\rm GeV} \leq
m_t \leq 200~{\rm GeV}$,  we find $1.87\times10^{-2} \leq \Delta r \leq
6.77\times10^{-2}$ \cite{sirlin}.   Using the experimental value given
by Langacker in Ref.~\cite{langacker}, we obtain
\begin{equation}
\label{precision}
\left[1-|{\cal O}_{e4}|^2\right]\left[1-|{\cal O}_{\mu 4}|^2\right]
\geq (0.9436)^2 \ ,
\end{equation}
at $90\%$ C.L., leading to an upper bound for $|{\cal O}_{e4}|^2|{\cal
O}_{\mu 4}|^2$ given by
\begin{equation}
\label{bound}
|{\cal O}_{e4}|^2|{\cal O}_{\mu 4}|^2 \leq 3.18\times10^{-3}\ ,
\end{equation}
or ${s_3}^2 \leq 0.11$ which is much less stringent then using the neutrino
counting in $Z$ decay as given in Eq.~(\ref{s3bound}).
In other words, the presence of a right-handed neutrino does not play an
important role for the precision measurement of $m_W/m_Z$.

The presence of a right-handed neutrino does violate the $\mu-e$
universality in charged current processes.  Let us first consider
the classic violation of the generation universality test in pion decay.
The ratio of the decay rates $R=\Gamma(\pi\rightarrow e
\nu)/\Gamma(\pi\rightarrow \mu \nu)$ in the presence of neutrino mixings
is given by
\begin{equation}
R=\frac{\Gamma(\pi\rightarrow e \nu)}{\Gamma(\pi\rightarrow \mu \nu)}
 = R_0 \frac{1-|{\cal O}_{e4}|^2}{1-|{\cal O}_{\mu 4}|^2} \ ,
\end{equation}
The experimental measurement relative to the standard model
expectation, $R/R_0$, is recently calculated to be $0.9969\pm0.0031\pm0.004$
\cite{marciano}, yielding
\begin{equation}
\label{emuconstraint}
\frac{1-|{\cal O}_{e4}|^2}{1-|{\cal O}_{\mu 4}|^2}=0.9969\pm0.0051.
\end{equation}

Next, we consider the charged current processes involving quarks,
where the CKM matrix (V) relevant
for nuclear $\beta$- and $K_{l3}$-decays are now modified by
\begin{equation}
\label{newV}
{\tilde V_{ud}} = \sqrt{\frac{1-|{\cal O}_{e4}|^2}
                        {1-|{\cal O}_{\mu 4}|^2}}V_{ud};\
{\tilde V_{us}} = \sqrt{\frac{1-|{\cal O}_{e4}|^2}
                      {1-|{\cal O}_{\mu 4}|^2}}V_{us}\ .
\end{equation}
Experimentally, the
quantity, $|{\tilde V_{ud}}|^2+|{\tilde V_{us}}|^2+|{\tilde V_{ub}}|^2$,
is measured to be $0.9979\pm0.0021$ \cite{databook}.
Since $|{\tilde V_{ub}}|^2 \leq
(0.01)^2$, its contribution is less than the uncertainty of the measurement.
Thus neglecting the contribution of $|{\tilde V_{ub}}|^2$ is well
justified.  Since the quark sector is not affected by the introduction
of singlet neutrinos, the unitarity of $V$ still holds, and exploiting
that gives
\begin{equation}
\label{Vconstraint}
\frac{1-|{\cal O}_{e4}|^2}{1-|{\cal O}_{\mu 4}|^2}=0.9974\pm0.0028 \ .
\end{equation}
Combining Eqs.~(\ref{emuconstraint}) and (\ref{Vconstraint}), we obtain
\begin{equation}
\label{universality}
0.9928 \leq \frac{1-|{\cal O}_{e4}|^2}{1-|{\cal O}_{\mu 4}|^2} \leq 1.0020
\end{equation}
at $90\%$ C.L..   In Fig. \ref{fig:Oconstraint}, we plot the constraints
obtained from Eqs.~(\ref{s3bound}) and (\ref{universality}).

As with the invisible decay width of $Z$, the presence of right-handed
neutrinos
will increase the lifetime of the $\tau$.  The updated world
averages of the tau lepton mass and lifetime are given by
$m_\tau=1770.0\pm0.4~\rm MeV$ and
$\tau_\tau=295.9\pm10^{-15}\rm~s$ respectively, and
the relevant leptonic branching ratios are
$B(\tau \rightarrow e \nu \nu)=17.77\pm0.15\%$ and
$B(\tau \rightarrow \mu \nu \nu)=17.48\pm0.18\%$,
as discussed in Ref.~\cite{weinstein}.
Using these values
for $m_\tau$ and $\tau_\tau$, the theoretical expectation for the
branching ratios are
$B(\tau \rightarrow e \nu \nu)|_{theor}=18.13\pm0.20\%$ and
$B(\tau \rightarrow \mu \nu \nu)|_{theor}=17.63\pm0.20\%$ \cite{marciano92}.
We can see that the experimental values for the branching ratios are
smaller than the theoretical expectation.  If the right-handed neutrino
is responsible for the discrepancies, we obtain
\begin{eqnarray}
\label{taudecay}
\frac{B(\tau\rightarrow e\nu\nu)}{B(\tau\rightarrow e\nu\nu)|_{theor}}+
\frac{B(\tau\rightarrow \mu\nu\nu)}{B(\tau\rightarrow \mu\nu\nu)|_{theor}}
&&=\left[1-|{\cal O}_{\tau 4}|^2\right]\left[2-|{\cal O}_{e 4}|^2 -
|{\cal O}_{\mu 4}|^2 \right] \nonumber \\
&&=1.9716\pm0.0204 \ .
\end{eqnarray}
At $90 \%$ C.L., this translates into limits on $c_2$ and $s_3$ as the
following,
\begin{equation}
s_3^2 ( 1+c_2^2 ) \leq 6.1\times10^{-2}\ ,
\end{equation}
or $s_3^2 \leq 6.1\times10^{-2}$ which is again less stringent than
Eq.~(\ref{s3bound}).

\section{lepton number violating (LNV) processes}
\label{sec:rare}

In this section, we compute, in the Feynman gauge, rare LNV decay
processes of the muon to one-loop accuracy.
To a very good approximation, we may take the masses of
$\nu_1$, $\nu_2$ and $\nu_3$ to be zero.
The rare processes we are interested are $\mu
\rightarrow e \gamma$, $\mu \rightarrow 3 e$, $e-\mu$ conversion in
nuclei, and neutrinoless double beta decay, $(2\beta)_{0\nu}$.
Details of the calculation of one-loop diagrams are given in
appendix~\ref{app:one-loop}.

Before going into the rare decay processes, we first consider the
large $m_4$ behavior of LNV penguin diagrams, $\mu-e-Z$ and $\mu-e-\gamma$,
and the two-W box diagrams given in
Fig~\ref{fig:FP}, \ref{fig:ZP} and \ref{fig:Bmu3e}.
Generic properties of decoupling effect for the see-saw model has
been considered recently in Ref.~\cite{cheng}.
Here, we treat $s_3$ as an independent parameter,
and consider the asymptotic behavior of the lepton flavor violating
effective vertices as $m_4$ goes to infinity.

Let us begin with the photon penguin diagrams shown in
Fig.~\ref{fig:FP}.  For large $m_4$, the effective vertices of photonic
penguin diagram, Eqs.~(\ref{F1}) and (\ref{F2}), are given by
\begin{eqnarray}
\label{F1limit}
\lim_{x_4 \to \infty} F_1 &=& s_1c_1s_2^2s_3^2
         \left(-\frac{\ln{x_4}}{6}\right)   \ , \\
\lim_{x_4 \to \infty} F_2 &=& s_1c_1s_2^2s_3^2
         \left(-\frac{1}{2}\right)  \ .
\label{F2limit}
\end{eqnarray}
The decoupling theorem is violated for both $F_1$ and $F_2$.

For the $Z$ penguin diagrams shown in Fig.~\ref{fig:ZP}, the effective vertex
given by Eq.~(\ref{ZPtot2}) for $x_4$ large becomes
\begin{equation}
\label{PZlimit}
\lim_{x_4 \to \infty} P_Z =  s_1c_1s_2^2s_3^2
   \left(s_3^2\frac{x_4}{4} \ln{x_4} \right) \ ,
\end{equation}
where this term comes from the Majorana nature of $\nu_4$.
Hence, we can see that the
decoupling theorem is also violated for the $Z$ penguin.

Finally, we consider the box diagram for $\mu \rightarrow 3 e$, shown in
Fig.~\ref{fig:Bmu3e}.
When $x_4$ is large, $B_{\mu \rightarrow 3 e}$ from Eq.~(\ref{Bmueee})
becomes
\begin{equation}
\label{Blimit}
\lim_{x_4 \to \infty} B_{\mu \rightarrow 3 e} = s_1c_1s_2^2s_3^2
  \left(s_1^2s_2^2s_3^2\frac{x_4}{2}\ln{x_4}\right) \ ,
\end{equation}
where this term comes from the diagrams in Fig.~\ref{fig:Bmu3e}(e,f,g,h).
Again, the decoupling theorem is violated.

We now consider each of these processes in some detail.

{\parindent=0pt a. {$\underline{\mu \rightarrow e \gamma}$}}.  The
transition amplitude for the process $\mu \rightarrow e \gamma$ is given
by
\begin{equation}
Amp(\mu \rightarrow e \gamma) = \frac{g^2e}{32\pi^2m_W^2}F_2~
     \epsilon^\mu(q)~{\overline e}~i\sigma_{\mu\nu}q^\nu m_\mu
    \frac{1+\gamma_5}{2}\ \mu \ ,
\end{equation}
where $\epsilon^\mu(q)$ is the polarization vector of the photon
with outgoing momentum $q$.  Hence, the decay branching ratio is given
by
\begin{equation}
B(\mu \rightarrow e \gamma)=\frac{3\alpha}{2\pi}|F_2|^2 \ .
\end{equation}

{\parindent=0pt b. {$\underline{\mu \rightarrow 3e}$}}.  This process
involves photon and $Z$ penguin as well as box diagrams.  The
interaction Lagrangian is given by
\begin{eqnarray}
{\cal L}(\mu \rightarrow 3e) =&& - \frac{\alpha G_F}{\sqrt{2}\pi} \left\{
  F_2~{\overline e}\ \gamma^\mu\ e ~{\overline e}~
    i\frac{\sigma_{\mu\nu}q^\nu}{q^2} m_\mu \frac{1+\gamma_5}{2}\ \mu
\right. \nonumber \\
 &&+\left. {\overline e}\ \gamma^\mu \left[ L \frac{1-\gamma_5}{2} +
       R \frac{1+\gamma_5}{2} \right] e~
    {\overline e}\ \gamma_\mu \frac{1-\gamma_5}{2}\ \mu \right\} \ ,
\end{eqnarray}
where $L$ and $R$ are defined as
\begin{eqnarray}
L &=& F_1+\frac{1}{s_W^2}\left( -\frac{1}{2}+s_W^2 \right) P_Z
   - \frac{1}{2s_W^2} B_{\mu \rightarrow 3e} \ , \\
R &=& F_1+ P_Z \ .
\end{eqnarray}
Hence, we obtain the branching ratio which is given by
\begin{equation}
B(\mu \rightarrow 3e)=\frac{\alpha^2}{16\pi^2}\left\{
   R^2+2L^2-4F_2(R+2L)+
  4F_2^2\left(4\ln\frac{m_\mu}{2m_e}-\frac{13}{6}\right) \right\} \ .
\end{equation}

{\parindent=0pt c. {\underline{$\mu-e$ conversion in nuclei}}}.  The
Feynman diagrams for this process can be obtained from that of $\mu
\rightarrow 3e$ by replacing the electron lines by quark lines.  Hence,
the interaction Lagrangian is given by
\begin{eqnarray}
{\cal L}(\mu-e) &=& - \frac{\alpha G_F}{\sqrt{2}\pi} \left\{
{\overline e}~i\frac{\sigma_{\mu\nu}q^\nu}{q^2}\ m_\mu\frac{1+\gamma_5}{2}\ \mu
  \left[ -\frac{2}{3}F_2~{\overline u}\ \gamma^\mu\ u +
        \frac{1}{3}F_2~{\overline d}\ \gamma^\mu\ d \right] \right. \nonumber
\\
  &+& \left. {\overline e}\ \gamma_\mu \frac{1-\gamma_5}{2}\ \mu
   \sum_{q=u,d} V_q~{\overline q}\ \gamma_\mu\ q~ \right\} \ ,
\end{eqnarray}
where the $V_q$'s are defined by
\begin{eqnarray}
V_u &=& -\frac{2}{3}F_1
   +\frac{1}{s_W^2} \left(\frac{1}{4}-\frac{2}{3}s_W^2 \right) P_Z
       - \frac{1}{4s_W^2} B^u_{\mu-e} \ , \\
V_d &=& \frac{1}{3}F_1
   +\frac{1}{s_W^2} \left(-\frac{1}{4}+\frac{1}{3}s_W^2 \right) P_Z
       - \frac{1}{4s_W^2} B^d_{\mu-e} \ .
\end{eqnarray}
In the above,  we include only the vector part of the quark current
because its contribution is larger than that of the axial part due to
the nuclear coherent effect \cite{goodman}.
Following the standard procedure \cite{feinberg,shanker,bernabeu}, we
obtain the transition rate for the $\mu-e$ conversion in nuclei as
follows:
\begin{equation}
\Gamma(\mu N \rightarrow e N)=\frac{\alpha^5 G_F^2 m_\mu^5}{16\pi^4}
      \frac{Z_{eff}^4}{Z} |F(-m_\mu^2)|^2 |Q_W|^2  \ ,
\end{equation}
with
\begin{equation}
Q_W=\left(\frac{2}{3}F_2+V_u\right)\left(2Z+N\right)
    +\left(-\frac{1}{3}F_2+V_d\right)\left(Z+2N\right) \,
\end{equation}
where $Z$ and $N$ are the atomic(or proton) and neutron numbers
for the nuclei,
and $|F(-m_\mu^2)|$  and $Z_{eff}$ are the nuclear form factor and the
effective atomic number.  For $^{48}_{22} Ti$, one has $|F(-m_\mu^2)|=0.54$
\cite{dreher} and $Z_{eff}=17.6$ \cite{ford}.

{}From the present data \cite{databook},  the branching ratios, $B(\mu
\rightarrow e \gamma)$, $B(\mu \rightarrow 3e)$ and $B(\mu\ Ti
\rightarrow e\ Ti)=\Gamma(\mu\ Ti \rightarrow e\ Ti)/\Gamma(\mu-capture)$ are
$4.9\times10^{-11}$, $1.0\times10^{-12}$ and $4.6\times10^{-12}$,
respectively, which translate into
\begin{eqnarray}
\label{limit1}
|F_2|^2 &&\leq 1.4\times10^{-8} \ , \\
\label{limit2}
\left[ R^2+2L^2-4F_2(R+2L)+65.0F_2^2 \right] &&\leq 3.0\times10^{-6}\ , \\
\label{limit3}
\left[ 70\left(\frac{2}{3}F_2+V_u\right)+
 74\left(-\frac{1}{3}F_2+V_d\right)\right]^2 &&\leq 2.6\times10^{-4} \ ,
\end{eqnarray}
respectively.  Note that the constraints, Eqs.~(\ref{limit1}),
(\ref{limit2}) and (\ref{limit3}), are independent of models.

At the first sight, it would seem that $\mu \rightarrow e \gamma$
provides the most stringent constraint among all three processes.
To compare among the experiments, let us consider the ratios
\begin{equation}
S_1~=~\frac{B(\mu \rightarrow 3e)}{B(\mu \rightarrow e \gamma)}\times
   \frac{4.9\times10^{-11}}{1.0\times10^{-12}} \ ,
\end{equation}
and
\begin{equation}
S_2~=~\frac{B(\mu\ Ti \rightarrow e\ Ti)}{B(\mu \rightarrow e \gamma)}\times
   \frac{4.9\times10^{-11}}{4.6\times10^{-12}} \ ,
\end{equation}
which provide a measure of the sensitivity of experiments.
For simplicity, we first neglect the contributions of the last terms in
Eqs.~(\ref{ZPtot2}) and (\ref{Bmueee}).  Hence the ratios
become independent of $\cal O$.   It can be easily shown that the ratios
$S_{1,2}$ are generically given by $(\ln x_4)^2$ for small and large
$x_4$ owing to the $Z$--penguin diagrams.
Thus, experiments $\mu \rightarrow 3e$ and $\mu-e$ conversion
have advantages over $\mu \rightarrow e \gamma$ in probing singlet
Majorana neutrinos.
We plot the ratios as functions of $m_4$ in Fig.~\ref{fig:expts}.
Furthermore,  $\mu-e$ conversion is further enhanced by the
coherence of the nuclei.  Therefore, we can use Eq.~(\ref{limit3}),
which is obtained from $\mu-e$ conversion in nuclei, to
place an upper bound on the mass of $\nu_4$ as a function of $|{\cal
O}_{\mu4}^\ast{\cal O}_{e4}|$.  In general, Eq.~(\ref{limit3}) depends on
$s_3$ and $|{\cal O}_{e4}|$.   Hence, we vary the values,
within the allowed range given in Fig.~\ref{fig:s3},
to obtain stronger and weaker bounds on $m_4$.
The result is depicted in Fig.~\ref{fig:xbound}.  In particular, for
$m_4 \geq m_W$, the stronger bound is given by the maximally allowed
value of $s_3$ whereas the weaker bound is given by $s_3=0$.

\section{neutrinoless double beta decay}
\label{sec:nonudecay}

The classic process to test for the
Majorana nature of neutrino masses is
neutrinoless
double beta decay, as depicted in Fig.~\ref{fig:bb0nu}.  The effective
Lagrangian is given by
\begin{equation}
\label{bb0nu}
{\cal L}_{\beta\beta 0\nu} = G_F^2~\frac{1}{q^2}~
  \left[\sum_{i=1}^{4}{\cal O}_{ei}^2 m_i \frac{q^2}{q^2-m_i^2}\right]
  {\overline e}\ \gamma^\mu\gamma^\nu (1+\gamma_5)\ e^c~
   {\overline u}\ \gamma_\mu (1-\gamma_5)\ d ~
     {\overline u}\ \gamma_\nu (1-\gamma_5)\ d \ ,
\end{equation}
where $q$ is the momentum carried by the internal neutrino line.
After integrating over all possible intermediate nuclear
states,  the quantity in the square bracket in Eq. (\ref{bb0nu})
becomes
\begin{equation}
\label{mnueff}
m_\nu(eff)=\sum_{i=1}^{4}{\cal O}_{ei}^2 m_i ~F(m_i,A) \ ,
\end{equation}
and \cite{halprin83}
\begin{equation}
\label{rho}
F(m,A)=<\frac{\exp^{-m r}}{r}>/<\frac{1}{r}> \ ,
\end{equation}
where $A$ is the total number of the nucleon.  Using the approximation
of uniform two-nucleon correlation of a hard core ($r_c = 0.5 \rm fm$)
\cite{halprin76}, Eq.~(\ref{rho}) becomes
\begin{equation}
F(m,A)=\frac{0.5}{(m R)^2} \left[ (1+m r_c)e^{-m r_c}-(1+2m R)e^{-2m R}
\right] \ ,
\end{equation}
where $R$ is the nuclear radius which is taken to be $R=1.2A^{1/3} \rm fm$.
Notice that if neutrinos
have opposite CP, there will be cancellation between their contributions.
In particular,
if both $m_3$ and $m_4$ are light, $F(m_{3,4},A)$ would be
approximately equal to unity.  Hence,
Eq.~(\ref{mnueff}) is then equal to  $s_1^2s_2^2(-c_3^2m_3+s_3^2m_4)$
which would be zero if we restrict ourselves to the see-saw mixing
relation, Eq.~(\ref{mixing}).  The cancellation is not complete when one
includes the nuclear correlation, Eq.~(\ref{rho}).  Again, here we consider
a general case where $s_3$ is considered as an independent parameter.

The best experimental limit on the quantity $|m_\nu(eff)|$ is
$1.5~\rm eV$ \cite{balysh}, which translates into
\begin{equation}
\label{limit4}
s_1^2s_2^2|-c_3^2~m_3 ~F(m_3,A)+s_3^2~m_4 ~F(m_4,A)| \leq 1.5~{\rm eV} \ .
\end{equation}
Let us first consider the contribution from
$\nu_4$.  Numerically, we have $s_3^2~m_4 ~F(m_4,A) \leq
1.6\times10^{-8}~(1.2\times10^{-10})~{\rm GeV}$
for $m_4=1~(2.5)~{\rm GeV}$, where $s_3$ is taken to be the maximally allowed
value shown in Fig.~\ref{fig:s3}.  Therefore, one would expect the
$\nu_3$ contribution to be important, leading to
\begin{equation}
\label{s1s2}
s_1^2s_2^2 \leq \frac{1.5 {\rm eV}}{m_3}.
\end{equation}
As a result, $s_1s_2$ would be very small when $\nu_3$ is relative heavy.
In particular, if $m_3~=~1~MeV$, we obtain $s_1s_2 < 10^{-3}$

\section{neutrino oscillations}
\label{sec:oscillation}

In the scenario we are considering, there are two massive and
two massless neutrinos, oscillation \cite{pentecorvo,maki}
of neutrino flavors will be allowed, leading to interesting phenomena
not available in the standard model.  When the mass of $\nu_4$ is greater
than the neutrino beam energy, there will be
three-flavor oscillation with one
oscillation wavelength, $\lambda_3=4\pi E/m_3^3$.  In addition, when we assume
$s_3$ to be very small, the oscillation mechanism depends only on two
mixings, namely $s_1$ and $s_2$.  Hence, for this situation the parameters
required to
describe neutrino oscillations are just $\lambda_3$, $s_1$ and $s_2$.

Let us first consider the neutrino-neutrino oscillation probabilities.
The oscillation probabilities corresponding to electron neutrino
$(\nu_e)$, which travels a distance $L$, are given by
\begin{eqnarray}
P(\nu_e \rightarrow \nu_e) &=& 1-4\sin^2(\frac{1}{2}k_3 L)
   \left(s_1^2s_2^2-s_1^4s_2^4\right) \,  \\
P(\nu_e \rightarrow \nu_\mu) &=& 4\sin^2(\frac{1}{2}k_3 L)
    \left(s_1^2s_2^4-s_1^4s_2^4\right) \, \\
P(\nu_e \rightarrow \nu_\tau) &=& 4\sin^2(\frac{1}{2}k_3 L)
    \left(s_1^2s_2^2-s_1^2s_2^4\right) \,
\end{eqnarray}
where
\begin{equation}
k_3=\frac{2\pi}{\lambda_3}=2.5m^{-1}\frac{m_3^2(eV)}{2E(MeV)} \ .
\end{equation}

When $m_3$ is in ${\rm MeV}$ range, the mixing, $s_2^2s_2^2$, is
constrained to be $10^{-6}$ or less, Eq.~(\ref{s1s2}).   Hence, the
oscillation becomes purely academic.  Furthermore, neutrino and
anti-neutrino oscillations, such as $\nu_e - {\overline{\nu_\tau}}$,
are also allowed.  However, it would be either suppressed by the ratio
$m_3/E$ or by small mixings, Eq.~(\ref{s1s2}), if $m_3 \gg O(1)~{\rm eV}$.

We next consider the case
when $m_3$ is small.  We will consider the following three cases:
\begin{itemize}
\item {{\underline{$k_3 L \ll 1$}}}:  When the neutrino
source is very close to the target, {\it i.e.} $L \ll 1/k_3$, the
oscillation effects are small. Hence, the probability $P(\nu_\mu
\rightarrow \nu_e)$ given by $|{\cal O}_{\mu4}|^2|{\cal O}_{e4}|^2$,
which is stringently constrained from $\mu-e$ conversion experiment,
would be in the order of $10^{-8}$.  Hence, lepton number violating
scatterings, such as $\nu_\mu N \rightarrow e N'$, are negligible.
\item{{\underline{$k_3 L \sim 1$}}}: In this case, there will be
oscillations.  In particular, the recent accelerator experment
\cite{angelini} allows us to probe $m_3$ in the range $0.1$ to $10~{\rm eV}$.
For atmospheric neutrino experiments, the mass range of $10^{-3} -
10^{-1}~{\rm eV}$ would be probed.  Constraints on three neutrino
mixings from atmospheric and reactor data has be studied
\cite{pantaleone} for $m_1=m_2=0$ and $m_3 > 0$.
\item{{\underline{$k_3 L \gg 1$}}}: In this case, the
oscillation effect is averaged out, namely $<\sin^2(\frac{1}{2}k_3 L)> =
1/2$.  In particular, the recent Gallex experiment, $P(\nu_e
\rightarrow \nu_e) = 0.66 \pm 0.12$, limits $s_1^2s_2^2$ to be within
either in the region of
$0.64 \leq s_1^2s_2^2 \leq 0.87$ or $0.13 \leq s_1^2s_2^2 \leq 0.36$ at
$ 1 \sigma$ level.
\end{itemize}

Therefore, even if $s_3$ turns out to be very small, neutrinoless double
beta decay and neutrino oscillation experiments provide another important
information for this model.

\section{conclusion}
\label{sec:con}

In this paper, we have studied the phenomenology of having right-handed
neutrino isosinglets.  In principle, when there are more than one
right-handed neutrinos, the masses of the light neutrinos
are not necessary given by $m_D^2/M$, where $m_D$ and $M$ are the Dirac
and Majorana mass terms.  In this paper, we regard the right-handed
neutrino $\nu^c$ as an effective collection of arbitrary number of
neutrinos and allow the mixing $s_3$ to be an independent parameter rather
than restricted by the see-saw relationships.

In the presence of $\nu^c$, neutrino flavor-changing $Z$ coupling exists
at tree level.
When $\nu_4$ is lighter than $Z$, the decay $Z \rightarrow \nu_3~\nu_4$ and
$Z \rightarrow \nu_4~\nu_4$ are allowed.  Hence, the decays of $\nu_4$
would give rise to exotic $Z$ decays, such as $Z \rightarrow e~\mu + X$.
Including the recent search for the lepton flavor violation in $Z$ decay, we
plot the result in Fig.~\ref{fig:s3}.  Furthermore, the violations of
universalities in charged current processes are also considered,
and the constraints on ${\cal O}_{\mu4}$ and ${\cal O}_{e4}$ are depicted in
Fig.~\ref{fig:Oconstraint}.   $\tau$ decays do not provide
 stringent constraints in this context.

Owing to the mixing and explicit Majorana mass term for $\nu^c$,
both separate and total lepton numbers are not conserved.  This
allows rare muon decays and neutrinoless double beta decays.  Among
various rare muon decay processes, $\mu-e$ conversion in nuclei places
the most severe constraints on the model.
Including the constraints derived from $Z$ decays,
we plot the upper bounds of $m_4$ as a function
of $|{\cal O}_{\mu 4}^\ast {\cal O}_{e 4}|$ in Fig.~\ref{fig:xbound}.

For the neutrinoless double beta decay, the contribution coming from
$\nu_3$ is more important, leading to the constraint $s_1^2s_2^2 \leq 1.5~
{\rm eV} / m_3$. In this model, three flavor oscillation
depends only on one oscillation wavelength and two mixing angles.
Thus, constraints from neutrino double beta decay as well as
the solar and atmospheric
neutrino experiments provide another important information for the model.

\begin{acknowledgments}

We thank J. Bernabeu for discussions.
This work was supported in part by the Department of Energy
under Grant No.\ DE-FG05-92ER40709, and by the
Engineering Research Council of Canada.

\end{acknowledgments}

\newpage

\begin{appendix}
\section{conditions obviating the see-saw mixing relationship}
\label{example}

The most general mass matrix for $n$-generations of left-handed and $m$
generations of right-handed neutrinos takes the form:
\begin{equation}
\pmatrix{0&\ldots&0&x_{11}&\ldots&x_{1m}\cr
\vdots&\ddots&\vdots&\vdots&\ddots&\vdots\cr
0&\ldots&0&x_{n1}&\ldots&x_{nm}\cr
x_{11}&\ldots&x_{n1}&M_{11}&\ldots&M_{1m}\cr
\vdots&\ddots&\vdots&\vdots&\ddots&\vdots\cr
x_{1m}&\ldots&x_{nm}&M_{m1}&\ldots&M_{mm}\cr
}
\end{equation}
We restrict our attention to the case $n\geq m$,
and assume that there is only an isodoublet Higgs field so that
Majorana masses for left-handed neutrinos are zero. The quantities $x_{ia}$
are Dirac masses, and are given by the product of the Yukawa coupling
constants and the vacuum expectation value of the isodoublet Higgs
field.    The $m\times m$ matrix $M_{ij}$ is the Majorana masses for the
right-handed neutrinos.

Without loss of generality, we may assume that $M_{ij}$
matrix is diagonal.     Next, we regard the vectors,
$\vec{x}_{i}$ with $i=1,...,m$, vectors in $n$-dimensional flavor space,
and subject the neutrinos to a rotation in this
space.   By this means, it will be possible to reduce $n-m$ of these
vectors to a form where their first $n-m$ components are zero.
Hence, $n-m$ neutrinos will be decoupled from massive neutrinos,
and remain massless at tree level.

For example, take the case of three generations of left-handed and two
generations of right-handed neutrinos. The Yukawa couplings define
for us two vectors in three dimensional space.  By the above argument, we
can project out the massless neutrino, leading to the following resultant
mass matrix:
\begin{equation}
\pmatrix{0&0&x_{21}&x_{22}\cr0&0&x_{31}&x_{32}\cr x_{21}&x_{31}&M_1&0
      \cr x_{22}&x_{32}&0&M_2}\; .
\end{equation}
The determinant of this mass matrix is given by $(\vec{x}_1 \times
\vec{x}_2)^2$.  For large $M_{1,2}$, and generic values for $\vec{x}_{1,2}$,
there will be two light and two heavy neutrinos.
The masses of the light neutrinos are given by the see-saw mass
relationships of the form $m_\nu \sim \vec{x}_1 \times \vec{x}_2/M$,
when $M$ is the collective mass for $M_{1,2}$.
In addition, the mixings of the light and heavy neutrino are of
order $x/M = \sqrt{m_\nu /M}$, where $x$ is a generic component of
$\vec{x}_{1,2}$.
{}From solar and atmospheric neutrino experiments, $m_\nu$
is required to be of order $10^{-3}~eV$.  If we take $M$ to be
$10^{15}~(10^2)~GeV$, then from the see-saw mass relationship $x$ will
be of order $10^2~(10^{-5})~GeV$.  As a result, the mixing $x/M$
would be very small, leading to negligible exotic processes such as $\mu
\rightarrow e \gamma$, $\mu \rightarrow 3 e$ and $\mu-e$
conversion in nuclei.

To enhance the mixing, one must have to evade from the see-saw mass
relationships.  For example,
when $\vec{x}_1 \times \vec{x}_2 \sim 0$, one
of the two light neutrinos will become massless.
In addition, when $M_1 \vec{x}_{2}^2 + M_2\vec{x}_{1}^2 \sim 0$,
the remaining light neutrino will also become massless.
Now, the mixing,  which is still given as $x/M$, cannot be rewritten as
the ratio of light to heavy neutrino masses.
Such apparently geometric conditions could
be remnants of a symmetry manifested only at higher energies.
In the phenomenological analysis described in this paper, we consider
such possibilities by allowing the mixing to be an independent parameter.

\section{one-loop diagram calculation}
\label{app:one-loop}
{\parindent=0pt 1. {\underline{Photon Penguin}}}  The calculation is
identical to that of sequential lepton models.  The effective vertex of
diagrams shown in Fig.~\ref{fig:FP} is
given by
\begin{equation}
\label{photonP}
\Gamma^\gamma_\mu=\frac{g^2e}{32\pi^2m_W^2}\left[
   F_1~\left( q^2\gamma_\mu-\not \hspace{-2pt}q q_\mu \right)
             {\overline e}\ \frac{1-\gamma_5}{2}\ \mu
 + F_2~{\overline e}~i\sigma_{\mu\nu}q^\nu~m_\mu\ \frac{1+\gamma_5}{2}\ \mu
\right]
\end{equation}
where
\begin{eqnarray}
\label{F1}
F_1&=&{\cal O}_{\mu 4}^\ast {\cal O}_{e 4} \left[
  \frac{x_4(12+x_4-7x_4^2)}{12(x_4-1)^3}+\frac{x_4^2(-12+10x_4-x_4^2)}
{6(x_4-1)^4}\ln{x_4} \right] \ , \\
\label{F2}
F_2&=&{\cal O}_{\mu 4}^\ast {\cal O}_{e 4} \left[
  \frac{x_4(1-5x_4-2x_4^2)}{4(x_4-1)^3}+\frac{3x_4^3}
{2(x_4-1)^4}\ln{x_4} \right] \ ,
\end{eqnarray}
where $x_4=m_4^2/m_W^2$.  It can be easily checked that the contribution
of $x_4$ is much larger than that of $x_3$ for $x_3 \ll 1$ and $x_3 \ll
x_4$.  For $x_3,~x_4 \ll 1$, the muon number violating processes would be
too small to be experimentally interested.
Hence, within the parameter space we are considering in this paper,
we can simply neglect the contribution of $x_3$.

{\parindent=0pt 2. \underline{$Z$ Penguin}}  $Z$ penguin diagrams depicted
in Fig.~\ref{fig:ZP} are more
complicated that the photon penguin because of the flavor changing
coupling, see Eqs.~(\ref{Znn}) and (\ref{Znn1}), as well as the Majorana
nature of neutrinos.   The $Z {\overline e} \mu$
effective vertex is defined as
\begin{equation}
\Gamma^Z_\mu=\frac{g^3}{32\pi^2\cos\theta_W}\sum_{\alpha}\Gamma_\alpha~
 {\overline e}\ \gamma_\mu\frac{1-\gamma_5}{2}\ \mu \ ,
\end{equation}
where the calculation of each diagram is given by
\begin{eqnarray}
\label{Z1a}
\Gamma_{a}
   &&=\sum_{i=1}^4 {\cal O}_{\mu i}^\ast {\cal O}_{e i} \left[
  \frac{1}{\varepsilon}-\frac{3}{4}+\frac{1}{2}F(x_i,x_i)+x_iG(x_i,x_i)
  \right] \nonumber \\
  &&-\sum_{i,j=1}^4{\cal O}_{\mu j}^\ast{\cal O}_{e i}
     {\cal O}_{R i}^\ast{\cal O}_{R j} \left[
  \frac{1}{\varepsilon}-\frac{3}{4}+\frac{1}{2}F(x_i,x_j)\right] \nonumber \\
 &&-\sum_{i,j=1}^4{\cal O}_{\mu j}^\ast{\cal O}_{e i}
     {\cal O}_{R i}{\cal O}_{R j}^\ast \left[\sqrt{x_ix_j}G(x_i,x_j)\right]
\;,\\
\label{Z1b}
\Gamma_{b}
   &&=\sum_{i=1}^4 {\cal O}_{\mu i}^\ast {\cal O}_{e i}(1-s_W^2)
   \left[-\frac{6}{\varepsilon}+\frac{1}{2}-3F(x_i,x_j)\right]\;,\\
\label{Z1cd}
\Gamma_{c+d}&&=\sum_{i=1}^4 {\cal O}_{\mu i}^\ast {\cal O}_{e i}s_W^2
  \left[2x_iG(1,x_i)\right]\;,\\
\label{Z1ef}
\Gamma_{e+f}&&=\sum_{i=1}^4 {\cal O}_{\mu i}^\ast {\cal O}_{e i}
 (\frac{1}{2}-s_W^2)\left[F(1,x_i)\right] \;,\\
\label{Z1g}
\Gamma_{g}
   &&=\sum_{i=1}^4 {\cal O}_{\mu i}^\ast {\cal O}_{e i} \left[
  -\frac{1}{2}x_iG(x_i,x_i)-\frac{1}{4}x_iF(x_i,x_i)\right] \nonumber \\
   &&+\sum_{i,j=1}^4{\cal O}_{\mu j}^\ast{\cal O}_{e i}
    {\cal O}_{R i}^\ast{\cal O}_{R j} \left[\frac{1}{2}x_ix_jG(x_i,x_j)\right]
\nonumber \\
     &&+\sum_{i,j=1}^4{\cal O}_{\mu j}^\ast{\cal O}_{e i}
     {\cal O}_{R i}{\cal O}_{R j}^\ast
    \left[\frac{1}{4}\sqrt{x_ix_j}F(x_i,x_j)\right] \;, \\
\label{Z1h}
\Gamma_{h}&&=\sum_{i=1}^4 {\cal O}_{\mu i}^\ast {\cal O}_{e i}
    (\frac{1}{2}-s_W^2)x_i\left[-\frac{1}{\varepsilon}-\frac{1}{4}
      -\frac{1}{2}x_iG(1,x_i) \right] \;, \\
\label{Z1ij}
\Gamma_{i+j}&&=-\Gamma_{h}
\end{eqnarray}
with
\begin{eqnarray}
F(a,b)&=&1-\frac{a^2}{(a-1)(a-b)}\ln{a}-\frac{b^2}{(b-1)(b-a)}\ln{b} \ , \\
G(a,b)&=&-\frac{a}{(a-1)(a-b)}\ln{a}-\frac{b}{(b-1)(b-a)}\ln{b} \ .
\end{eqnarray}
Each of the divergent diagrams in Figs.~\ref{fig:ZP}(a-f)
is finite after summing
all the internal neutrinos due to the unitarity of ${\cal O}$ and
Eq.~(\ref{diagonal});
the divergences cancel among diagrams in Fig.~\ref{fig:ZP}(g-j).
Note that  dependence on $s_W^2$ disappears when all the diagrams
are summed because of gauge invariance.
The last terms in the first lines of Eqs.~(\ref{Z1a}) and (\ref{Z1g})
are due to Majorana property of neutrinos and the last two lines in
Eqs.~(\ref{Z1a}) and (\ref{Z1g}) are due to flavor changing $Z$ couplings.
Summing over all contributions, we obtain
\begin{eqnarray}
\label{PZ}
\sum_{\alpha}\Gamma_\alpha&&=\sum_{i=3}^4{\cal O}_{\mu i}^\ast {\cal
                               O}_{e i}\left[ \left(
  \frac{x_i^2-6x_i}{2(x_i-1)}+\frac{3x_i^2+2x_i}{2(x_i-1)^2}\ln{x_i}\right)
  +\left(-\frac{3x_i}{4(x_i-1)}
        +\frac{x_i^3-2x_i^2+4x_i}{4(x_i-1)^2}\ln{x_i} \right)
 \right] \nonumber \\
&&+\sum_{i,j=3}^4{\cal O}_{\mu j}^\ast{\cal O}_{e i}
    \left\{ {\cal O}_{R i}^\ast{\cal O}_{R j}\left[
  -\frac{x_ix_j}{2(x_i-x_j)}\ln{x_i}-\frac{x_jx_i}{2(x_j-x_i)}\ln{x_j}
   \right] \right. \nonumber \\
&&+\left. {\cal O}_{R i}{\cal O}_{R j}^\ast \sqrt{x_ix_j} \left[
   \frac{1}{4}+\frac{4x_i-x_i^2}{4(x_i-1)(x_i-x_j)}\ln{x_i}
   +\frac{4x_j-x_j^2}{4(x_j-1)(x_j-x_i)}\ln{x_j} \right]\right\}\;,
\end{eqnarray}
Note that
$|{\cal O}_{\mu i}^\ast {\cal O}_{R i}|
=|{\cal O}_{\mu i}^\ast {\cal O}_{R i}^\ast|=c_1s_2c_3s_3$ and
$|{\cal O}_{e i}^\ast {\cal O}_{R i}|
=|{\cal O}_{e i} {\cal O}_{R i}| = s_1s_2c_3s_3$, for $i=3,4$.
Hence, flavor changing $Z$ coupling contributions are also dominated by
$x_4$, and Eq.~(\ref{PZ}), to a very good approximation, becomes
\begin{eqnarray}
P_Z &&=\sum_{\alpha}\Gamma_\alpha \nonumber \\
    &&={\cal O}_{\mu 4}^\ast{\cal O}_{e 4}\left[ \left(
\frac{x_4^2-6x_4}{2(x_4-1)}+\frac{3x_4^2+2x_4}{2(x_4-1)^2}\ln{x_4}\right)
  +\left(-\frac{3x_4}{4(x_4-1)}
        +\frac{x_4^3-2x_4^2+4x_4}{4(x_4-1)^2}\ln{x_4} \right)
 \right. \nonumber \\
&&+|{\cal O}_{R 4}|^2\left. \left(
   \frac{-2x_4^2+5x_4}{4(x_4-1)}+\frac{-x_4^3+2x_4^2-4x_4}
     {4(x_4-1)^2}\ln{x_4} \right) \right] \ ,
\label{ZPtot}
\end{eqnarray}
where the parenthesis help us to identify various contributions.  We can
also rewrite Eq.~(\ref{ZPtot}) as
\begin{equation}
\label{ZPtot2}
P_Z={\cal O}_{\mu 4}^\ast{\cal O}_{e 4}\left[ \left(
-\frac{5x_4}{2(x_4-1)}+\frac{2x_4+3x_4^2}{2(x_4-1)^2}\ln{x_4} \right)
+s_3^2 \left(\frac{2x_4^2-5x_4}{4(x_4-1)}+\frac{x_4^3-2x_4^2+4x_4}
     {4(x_4-1)^2}\ln{x_4} \right) \right] .
\end{equation}

{\parindent=0pt 3.{\underline {$\mu \rightarrow 3e$ box diagrams}}}
There are two
different classes of box diagrams, Fig.~\ref{fig:Bmu3e}(a,b,c,d) and
\ref{fig:Bmu3e}(e,f,g,h), which contribute to the decay of
$\mu \rightarrow 3e$.  The effective interaction Lagrangian is defined as
\begin{equation}
\label{B3e}
\frac{g^4}{32\pi^2m_W^2} \sum_\alpha B_\alpha~{\overline e}\ \gamma_\mu
 \frac{1-\gamma_5}{2}\ e~{\overline e}\ \gamma^\mu\frac{1-\gamma_5}{2}\ \mu\ .
\end{equation}
The calculation of each diagram is given by
\begin{eqnarray}
\label{Ba}
B_{a}&=&{\cal O}_{\mu 4}^\ast {\cal O}_{e 4}\left[
 \left(\frac{x_4}{x_4-1}-\frac{x_4}{(x_4-1)^2}\ln{x_4}\right)
  +|{\cal O}_{e 4}|^2 \left(\frac{x_4^2+x_4}{(x_4-1)^2}
      +\frac{2x_4^2}{(x_4-1)^3}\ln{x_4}\right) \right] \ , \\
\label{Bb}
B_{b}&=&{\cal O}_{\mu 4}^\ast {\cal O}_{e 4} |{\cal O}_{e 4}|^2
  \left[-\frac{x_4^3+x_4^2}{4(x_4-1)^2}+\frac{x_4^3}{2(x_4-1)^3}\ln{x_4}
     \right] \ , \\
\label{Bcd}
B_{c+d}&=&{\cal O}_{\mu 4}^\ast {\cal O}_{e 4} |{\cal O}_{e 4}|^2
  \left[\frac{4x_4^2}{(x_4-1)^2}+2\frac{x_4^3+x_4^2}{(x_4-1)^3}\ln{x_4}
      \right]\ , \\
\label{Be}
B_{e}&=&{\cal O}_{\mu 4}^\ast {\cal O}_{e 4} |{\cal O}_{e 4}|^2
  \left[-\frac{4x_4}{(x_4-1)^2}+2\frac{x_4^2+x_4}{(x_4-1)^3}\ln{x_4}
      \right] \ , \\
\label{Bf}
B_{f}&=&{\cal O}_{\mu 4}^\ast {\cal O}_{e 4} |{\cal O}_{e 4}|^2
   \left[ -\frac{x_4^3}{(x_4-1)^2}+\frac{x_4^4+x_4^3}{2(x_4-1)^3} \ln{x_4}
      \right] \ , \\
\label{Bgh}
B_{g+h}&=&{\cal O}_{\mu 4}^\ast {\cal O}_{e 4} |{\cal O}_{e 4}|^2
  \left[\frac{x_4^2+x_4}{(x_4-1)^2}-\frac{2x_4^2}{(x_4-1)^3}\ln{x_4}
      \right] \ ,
\end{eqnarray}
where the contributions from $x_3$ is negligible.  Again
$B_{a,b,c,d}$ are the same as the
sequential lepton models, and
$B_{e,f,g,h}$ are due to the Majorana
properties of neutrinos.
Summing up all the contributions, Eqs.~(\ref{Ba}-\ref{Bgh}), we obtain
\begin{eqnarray}
\label{Bmu3e}
B_{\mu \rightarrow 3e} &&= \sum_{\alpha} B_\alpha  \nonumber \\
&&= {\cal O}_{\mu 4}^\ast {\cal O}_{e 4} \left[
  \frac{x_4}{x_4-1}-\frac{x_4}{(x_4-1)^2}\ln{x_4}
+|{\cal O}_{e 4}|^2 \left(\frac{-4x_4+11x_4^2-x_4^3}{4(x-1)^2}
  -\frac{3x_4^3}{2(x_4-1)^3}\ln{x_4}\right) \right. \nonumber \\
&&+\left. |{\cal O}_{e 4}|^2 \left(\frac{-3x_4+x_4^2-x_4^3}{(x_4-1)^2}
  +\frac{4x+x^3+x^4}{2(x_4-1)^3}\ln{x_4} \right) \right] \ ,
\end{eqnarray}
or
\begin{eqnarray}
\label{Bmueee}
B_{\mu \rightarrow 3e}
&&= {\cal O}_{\mu 4}^\ast {\cal O}_{e 4} \left[
  \frac{x_4}{x_4-1}-\frac{x_4}{(x_4-1)^2}\ln{x_4} \right. \nonumber \\
  &&+\left. |{\cal O}_{e 4}|^2 \left(\frac{-16x_4+15x_4^2-5x_4^3}{4(x_4-1)^2}
    +\frac{4x-2x^3+x_4^4}{2(x_4-1)^3}\ln{x_4} \right) \right] \ .
\end{eqnarray}

{\parindent=0pt 4.{\underline {$\mu - e$ conversion box diagrams}}}
The box diagrams corresponding to $\mu - e$ conversion in nuclei can be
obtained from Figs.~\ref{fig:Bmu3e}(a,b,c,d) by replacing the electron lines
with
quark lines.  The effective interactions are defined as
\begin{equation}
\label{Bmueqq}
\frac{g^4}{32\pi^2m_W^2}{\overline e}\ \gamma_\mu \frac{1-\gamma_5}{2}\ \mu
  \left[ B^u_{\mu-e}~{\overline u}\ \gamma^\mu\frac{1-\gamma_5}{2}\ u +
     B^d_{\mu-e}~{\overline d}\ \gamma^\mu\frac{1-\gamma_5}{2}\ d \right] \ ,
\end{equation}
where $B^u_{\mu-e}$ and $B^d_{\mu-e}$ are given by
\begin{eqnarray}
\label{Bu}
B^u_{\mu-e}&=&{\cal O}_{\mu 4}^\ast {\cal O}_{e 4}
     \left[\frac{4x_4}{x_4-1}-\frac{4x_4}{(x_4-1)^2}\ln{x_4}\right]
\ , \\
\label{Bd}
B^d_{\mu-e}&=&{\cal O}_{\mu 4}^\ast {\cal O}_{e 4}
     \left[\frac{x_4}{x_4-1}-\frac{x_4}{(x_4-1)^2}\ln{x_4}\right] \ ,
\end{eqnarray}
and we have neglected the contribution from the top-quark
because $|{V^{CKM}}_{td}|^2(m_t^2/m_W^2) \ll 1$.

\end{appendix}

\begin{figure}
\caption{The $90\%$ C.L. upper bound of $s_3^2$ as a function of
$m_4$ obtained from the $Z$ decay.}
\label{fig:s3}
\end{figure}
\begin{figure}
\caption{The $90\%$ C.L. constraint of $|{\cal O}_{e4}|^2$ as a function
of $|{\cal O}_{\mu 4}^\ast {\cal O}_{e 4}|^2$ obtained from the
universality constraints (solid curves) and the upper bound of
$|{\cal O}_{\mu 4}^\ast {\cal O}_{e 4}|^2 \leq 1/4~s_3^4$ obtained in
Eq.~(3.2) (dashed vertical line).}
\label{fig:Oconstraint}
\end{figure}
{\parindent=0pt
\begin{figure}
\caption{Photon penguin diagrams for $\mu-e-\gamma$ vertex.}
\label{fig:FP}
\end{figure}}
\begin{figure}
\caption{$Z$ penguin diagrams for $\mu-e-Z$ vertex.}
\label{fig:ZP}
\end{figure}
\begin{figure}
\caption{Box diagrams for the process $\mu \rightarrow 3 e$.  The
crosses correspond to flipping the neutrino helicities.}
\label{fig:Bmu3e}
\end{figure}
\begin{figure}
\caption{Sensitivity $(S)$ of experiments, $\mu \rightarrow 3e$ (dashed
line) and $\mu-e$ conversion (solid line), relative to $\mu \rightarrow e
\gamma$, where $S~=~S_1$ and $S_2$ respectively.  }
\label{fig:expts}
\end{figure}
\begin{figure}
\caption{The stronger (solid line) and weaker (dashed line) upper bounds
on $m_4$ derived from $\mu-e$ conversion experiment
as a function of $|{\cal O}_{\mu 4}^\ast {\cal O}_{e 4}|$,
where we have included the bound obtained from $Z$-decays.}
\label{fig:xbound}
\end{figure}
\begin{figure}
\caption{The Mechanism for neutrinoless double beta decay, where the
cross corresponds to flipping the neutrino helicity.}
\label{fig:bb0nu}
\end{figure}

\end{document}